\documentclass{article}
\usepackage{spconf,amsmath,graphicx}
\usepackage{url}
\usepackage{multirow}
\usepackage{amssymb}
\usepackage{amsmath}
\usepackage{bm}

\title{comparison of user models based on GMM-UBM and i-vectors for speech, handwriting, and gait assessment of Parkinson's disease patients}
\name{J. C. Vasquez-Correa$^{1,2\star}$ \qquad T. Bocklet$^{3}$ \qquad J. R. Orozco-Arroyave$^{1,2}$ \qquad E. N\"oth$^{1}$}
\address{$^1$Pattern Recognition Lab. Friedrich-Alexander Universit\"at, Erlangen-N{\"u}rnberg, Germany\\
	$^2$ Faculty of Engineering. Universidad de Antioquia UdeA, Calle 70 No. 52-21, Medell\'in, Colombia \\
	$^3$ Technische Hochschule N\"urnberg, Germany \\
	$^\star$corresponding author: \url{juan.vasquez@fau.de}}

\begin{document}
%
\maketitle
\begin{abstract}

Parkinson's disease is a neurodegenerative disorder characterized by the presence of different motor impairments. Information from speech, handwriting, and gait signals have been considered to evaluate the neurological state of the patients. On the other hand, user models based on Gaussian mixture models - universal background models (GMM-UBM) and i-vectors are considered the state-of-the-art in biometric applications like speaker verification because they are able to model specific speaker traits. This study introduces the use of GMM-UBM and i-vectors to evaluate the neurological state of Parkinson's patients using information from speech, handwriting, and gait. The results show the importance of different feature sets from each type of signal in the assessment of the neurological state of the patients.

\end{abstract}
\section{Introduction}
\label{sec:intro}

Parkinson's disease (PD) is a neurological disorder characterized by the progressive loss of dopaminergic neurons in the midbrain,  producing several motor and non-motor impairments~\cite{Hornykiewicz1998}. PD affects all of the sub-systems involved in motor activities like speech production, walking, or handwriting. The severity of the motor symptoms is evaluated with the third section of the movement disorder society - unified Parkinson's disease rating scale (MDS-UPDRS-III)~\cite{Goetz2008}. 
The assessment requires the patient to be present at the clinic, which is expensive and time-consuming because several limitations, including the availability of neurologist and the reduced mobility of patients. 
The evaluation of motor symptoms is crucial for clinicians to make decisions about the medication or therapy for the patients~\cite{Patel2010}. The analysis of signals such as gait, handwriting, and speech helps to assess the motor symptoms of patients, providing  objective information to clinicians to make timely decisions about the treatment.

Several studies have analyzed different signals such as speech, gait, and handwriting to monitor the neurological state of the PD patients.
Speech was considered in~\cite{tu2017objective} to predict the MDS-UPDRS-III score of 61 PD patients using spectral and glottal features. The authors computed the Hausdorff distance between a speaker from the test set and the speakers in the training set. The neurological state of the patients was predicted with a Pearson's correlation of up to 0.58.
In~\cite{smith2017vocal} the authors predicted the MDS-UPDRS-III score of 35 PD patients  with features based on articulation and prosody analyses, and a  Gaussian staircase regression. The authors reported moderate Spearman's correlations ($\rho$=0.42).
Handwriting was considered in~\cite{mucha2018identification}, to predict the H\&Y score of 33 PD patients using kinematic features and a regression based on gradient-boosting trees. The  H\&Y score was predicted  with an equal error rate of  12.5\%.
Finally, regarding gait features, in~\cite{aghanavesi2019motion} the authors predicted a lower limbs subscore of the MDS-UPDRS-III from 19 PD patients, using several harmonic and non-linear features, and  a support vector regression (SVR) algorithm. The subscore for lower limbs was predicted with an intra-class correlation coefficient of 0.78.

According to the literature, most of the related works consider only one modality. Multimodal analyses, i.e., considering information from different sensors, have not been extensively studied.
In~\cite{barth2012combined} the authors combined information from 
statistical and spectral features extracted from 
handwriting and gait signals. The fusion of features improved the accuracy of the classification between PD and healthy control (HC) subjects.
Previous studies~\cite{vasquez2017gcca,vasquez2019multimodal} suggested that the combination of modalities also improved the accuracy in the prediction of the neurological state of the patients.
This study proposes the use of different features extracted from speech, handwriting, and gait to evaluate the neurological state of PD patients. The prediction is performed with user models based on Gaussian mixture models - universal background models (GMM-UBM) and i-vectors. To the best of our knowledge, this is one of the few studies for multimodal analysis of PD patients, and the first one that considers multimodal user models to evaluate the neurological state of the patients.

\section{Methods}
\label{sec:methods}

\begin{figure}[!ht]
    \centering
    \includegraphics[width=\linewidth]{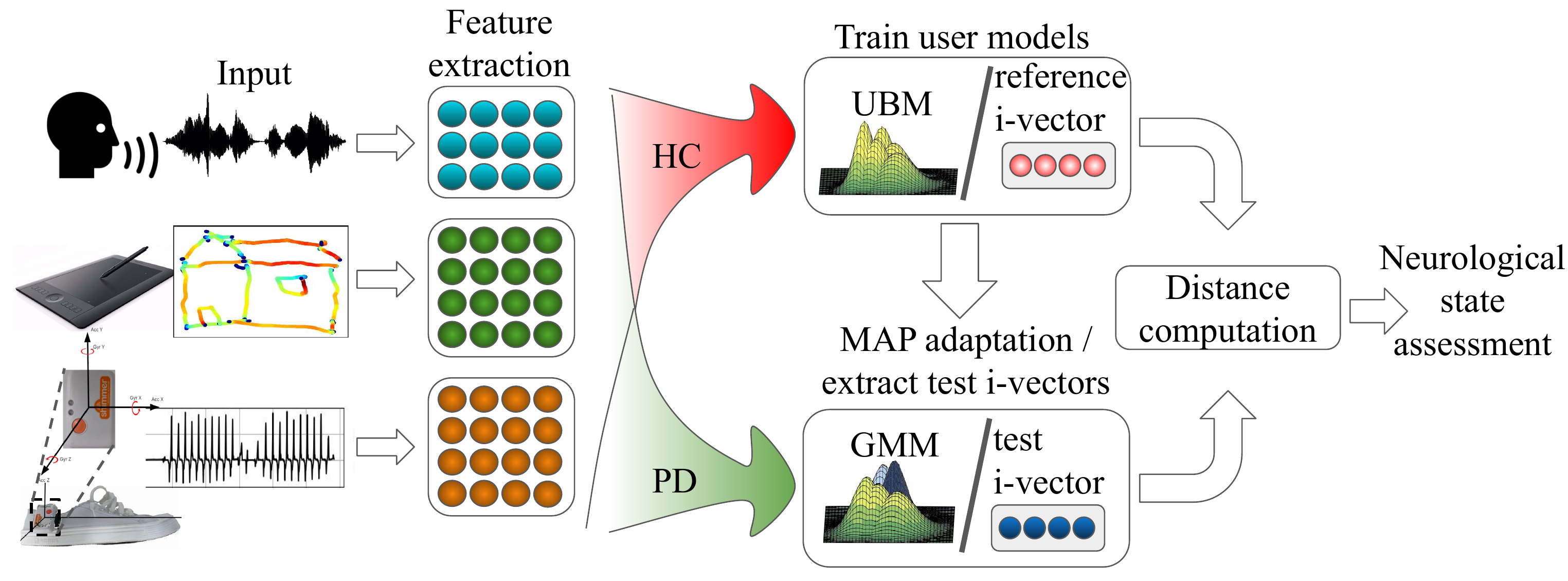}
    \caption{General methodology followed in this study.}
    \label{fig:method}
\end{figure}

The methods used in this study are summarized in Figure~\ref{fig:method}. Speech, handwriting, and gait signals are characterized using different feature extraction strategies. Then, data from HC subjects are used to train user models based on GMM-UBM and i-vector systems. 
For the case of the GMM-UBM, data from PD patients were used to adapt the UBMs into GMMs, creating a specific GMM for each patient. On the other hand, for the i-vector modeling, a reference i-vector was created with data from HC subjects with similar age and gender of the patients, thus i-vectors extracted from the patients can be compared with a personalized reference model. Finally, distance measures are computed between the reference models and those adapted/extracted from the PD patients. The computed distance is correlated with the neurological state of the patients based on the MDS-UPDRS-III scale.

\subsection{Speech features}

\textbf{Phonation:} these features model abnormal patterns in the vocal fold vibration. Phonation features are extracted from the voiced segments. The feature set includes descriptors computed for 40\,ms frames of speech, including jitter, shimmer, amplitude perturbation quotient, pitch perturbation quotient, the first and second derivatives of the fundamental frequency $F\raisebox{-.4ex}{\scriptsize 0}$, and the log-energy~\cite{orozco2018neurospeech}.

\textbf{Articulation:} these features model aspects related to the movements of limbs involved in the speech production. The features considered the energy content in onset segments~\cite{orozco2018neurospeech}. The onset detection is based on the computation of $F\raisebox{-.4ex}{\scriptsize 0}$. Once the border between unvoiced and voiced segments is detected, 40\,ms of the signal are taken to the left and to the right, forming a segment with 80\,ms length. The spectrum of the onset is distributed into 22 critical bands according to the Bark scale, and the Bark-band energies (BBE) are calculated. 12 MFCCs and their first two derivatives are also computed in the transitions to complete the feature set.

\textbf{Prosody: } for these features, the log-$F\raisebox{-.4ex}{\scriptsize 0}$ and the log-energy contours of the voiced segments were approximated using Lagrange polynomials of order $P=5$. A 13-dimensional feature vector is formed by concatenating the six coefficients computed from the log-$F\raisebox{-.4ex}{\scriptsize 0}$ and the log-energy contours, in addition to the duration of the voiced segment~\cite{Dehak2007}. The aim of these features is to model speech symptoms such as monotonicity ad mono-loudness in the patients.

\textbf{Phonological: }  these features are represented by a vector with interpretable information about the placement and manner of articulation. The different phonemes of the Spanish language are grouped into 18 phonological posteriors. The phonological posteriors were computed with a bank of parallel recurrent neural networks to estimate the probability of occurrence of a specific phonological class~\cite{vasquez2019phonet}. 

\subsection{Handwriting features}

Handwriting features are based on the trajectory of the strokes in vertical, horizontal, radial, and angular positions. We computed the velocity and acceleration of the strokes in the different axes, in addition to the pressure of the pen, the azimuth angle, the altitude angle, and their derivatives. Finally, we considered features based on the in-air movement before the participant put the pen on the tablet's surface. Additional information of the features can be found in~\cite{rios2019}.

\subsection{Gait features}

\textbf{Harmonic:} these features model the spectral wealth and the harmonic structure of the gait signals obtained from the inertial sensors. We computed the continuous wavelet transform with a Gaussian wavelet. The feature set is formed with the energy content in 8 frequency bands from the scalogram, three spectral centroids, the energy in the in the 1st, 2nd, and 3rd quartiles of the spectrum, the energy content in the locomotor band (0.5--3\,Hz), the energy content in the freeze band (3--8\,Hz), and the freeze index, which is the ratio between the energy in the locomotor and freeze bands~\cite{zach2015identifying,rezvanian2016towards}.

\textbf{Non-linear: } gait is a complex and non-linear activity that can be modeled with non-linear dynamics features. The first step to extract those features is the phase space reconstruction, according to the Taken's theorem. Different features can be extracted from the reconstructed phase space to assess the complexity and stability of the walking process. The extracted features include the correlation dimension, the largest Lyapunov exponent, the Hurst exponent, the detrended fluctuation analysis, the sample entropy, and the Lempel-Ziv complexity~\cite{perez2018non}.

\subsection{User models based on GMM-UBM}

GMM-UBM systems were proposed recently to quantify the disease progression of PD patients~\cite{arias2018speaker}. We propose to extend the idea to multimodal GMM-UBM systems. The main hypothesis is that speech, handwriting, or gait impairments of PD patients can be modeled by comparing a GMM adapted for a patient with a reference model created with recordings from HC subjects.
GMMs represent the distribution of feature vectors extracted from the different signals from a single PD patient. When the GMM is trained using features extracted from a large sample of subjects, the resulting model is a UBM. 
The model for each PD patient is derived from the UBM by adapting its parameters following a maximum a posteriori process. 
Then, the neurological state of the patients is estimated by comparing the adapted model with the UBM using a distance measure. We use the Bhattacharyya distance, which considers differences in the mean vectors  and covariances matrices between the UBM and the user model~\cite{you2010gmm}.

\subsection{User models based on i-vectors}

I-vectors are used to transform the original feature space into a low-dimensional representation called total variability space via joint factor analysis~\cite{Dehak2011}. For speech signals, such a space models the inter- and intra-speaker variability, in addition to channel effects. For this study we aim to capture changes in speech, handwriting, and gait due to the disease~\cite{garcia2018multimodal}.  I-vectors have been considered previously to model handwriting~\cite{christleinhandwriting} and gait data~\cite{san2017vector}.
Similar to the GMM-UBM systems, we train the i-vector extractor with data from HC subjects, and compute a reference i-vector to represent healthy speech, handwriting, or gait. Then, we extract i-vectors from PD patients, and compute the cosine distance between the patient i-vector and the reference.

\section{Data}
\label{sec:data}

We considered an extended version of the PC-GITA corpus~\cite{Orozco2014DB}. This version contains speech, handwriting, and gait signals, collected from 106 PD patients and 87 HC subjects. All of the subjects are Colombian Spanish native speakers. The patients were labeled according to the MDS-UPDRS-III scale. Table~\ref{tab:people} summarizes clinical and demographic aspects of the participants included in the corpus. 

\begin{table}[!ht]
\centering
\caption{Clinical and demographic information of the subjects. [F/M]: Female/Male. Average(Standard deviation). TD: Time since diagnosis. TD and age are given in years.}
\label{tab:people}
\resizebox{\linewidth}{!}{
\begin{tabular}{lcc}
\hline

\hline
\multicolumn{1}{c}{\textbf{}}  & PD patients          & HC subjects  \\ \hline
Gender {[}F/M{]}               & 49/57                 & 43/44                            \\
Age {[}F/M{]}                  & 60.9(11.2)/64.7(9.4)  & 61.4(9.8)/64.9(10.5)             \\
TD {[}F/M{]} & 15.5(14.5)/8.1(5.9)   &          --                          \\
MDS--UPDRS--III {[}F/M{]}      & 36.2(18.1)/36.3(18.9) &        --                              \\ 
\hline
\end{tabular}}

\end{table}

Speech signals were recorded with a sampling frequency of 16\,kHz and 16-bit resolution. The same speech tasks recorded in the PC-GITA corpus~\cite{Orozco2014DB}, except for the isolated words, are included in this extended version. 
Handwriting data consist of online drawings captured with a tablet Wacom cintiq 13-HD with a sampling frequency of 180\,Hz. The tablet captures six different signals: x-position, y-position, in-air movement, azimuth, altitude, and pressure. The subjects performed a total of 14 exercises divided into writing and drawing tasks. Additional information about the handwriting exercises can be found in~\cite{vasquez2019multimodal}.
Gait signals were captured with the eGaIT system, which consists of a 3D-accelerometer (range $\pm$6g) and a 3D gyroscope (range $\pm$500$^\circ$/s) attached to the external side (at the ankle level) of the shoes~\cite{barth2015stride}. Data from both feet were captured with a sampling frequency of 100\,Hz and 12-bit resolution. The exercises included 20 meters walking with a stop after 10 meters, 40 meters walking with a stop every 10 meters, \emph{heel-toe tapping}, and the \emph{ time up and go} test.

\section{Experiments and results}
\label{sec:results}

Data from HC subjects were used to train the UBMs and the i-vector extractors.  For the GMM-UBM system, data from PD patients were used to adapt the UBMs into GMMs. The Bhattacharya distance is used to compare the GMM and the UBM. For the i-vectors, a reference was created by averaging the i-vectors extracted from HC subjects that have same gender and similar age of the patients (in a range of $\pm$ 2 years). I-vectors extracted from PD patients are compared to the reference i-vector using the cosine distance. The computed distances are correlated with the MDS-UPDRS-III score of the patients.

\subsection{User models from different modalities}

The correlation between the neurological state of the patients and the user models based on GMM-UBM and i-vectors is shown in Table~\ref{tab:results1} for the  speech, handwriting, and gait features. The results indicate that for gait and speech signals, the user models based on GMM-UBM systems are more accurate than the i-vectors. This can be explained because the distance between the adapted GMM and the UBM considers more information about the statistical distribution of the population than for the case of i-vectors, where the reference for healthy subjects is reduced to a single vector. A  ``strong" correlation is obtained with the harmonic features (gait analysis) modeled with the GMM-UBM system ($\rho$=0.619). This is expected because most of the items used by the neurologist in the MDS-UPDRS-III are based on the movement of the lower limbs. For handwriting features, ``weak" correlations are obtained both with the GMM-UBM and the i-vector systems. The correlations obtained with speech features are not robust to model the general neurological state of the patients. This result can be explained because the MDS-UPDRS-III is a complete neurological scale that consider speech impairments in only one of the 33 items of the total scale~\cite{Goetz2008}. 

\begin{table}[htbp]
\caption{Correlation between the MDS-UPDRS-III scale of PD patients and the user models based on GMM-UBM and i-vectors extracted using different feature sets. \textbf{r:} Pearson's correlation coefficient, $\rho$: Spearman's correlation coefficient.}
\resizebox{\linewidth}{!}{
\begin{tabular}{llcccc}
\hline
\textbf{}               & \textbf{}         & \multicolumn{2}{c}{\textbf{GMM-UBM}}      & \multicolumn{2}{c}{\textbf{I-vector}}     \\
\textbf{Modality}       & \textbf{Features} & \textbf{r} & $\mathbf{\rho}$ & \textbf{r} & $\mathbf{\rho}$ \\ \hline
\multirow{2}{*}{Gait}   & Harmonic          & 0.436      & 0.619                        & 0.301      & 0.354                        \\
                        & Non-linear        & 0.262      & 0.312                        & 0.345      & 0.381                        \\ \hline
Handwriting             & Kinematic         & 0.339      & 0.261                        & 0.237      & 0.346                        \\ \hline
\multirow{4}{*}{Speech} & Phonation         & 0.176      & 0.198                        & 0.279      & 0.244                        \\
                        & Articulation      & 0.179      & 0.195                        & 0.225      & 0.213                        \\
                        & Prosody           & 0.067      & 0.202                        & 0.191      & 0.190                        \\
                        & Phonological      & 0.266      & 0.298                        & 0.212      & 0.197                        \\ \hline
\end{tabular}}
\label{tab:results1}
\end{table}

\subsection{Multimodal user models}

The user models extracted from all feature sets using the GMM-UBM system were combined by concatenating the distance between the user model and the UBM for each feature set. A linear regression was trained with the matrix of distances to predict the MDS-UPDRS-III scale. The model was trained following a leave one subject out cross-validation strategy. The results of the fusion are shown in Table~\ref{tab:results2}. The Spearman's correlation increases by 2.4\%, absolute  with respect to the one obtained only with the harmonic features. Additional regression algorithms based on random forest regression or SVRs were considered, however, they overfitted the test set, predicting only the mean value of the total scale.

\begin{table}[!ht]
\caption{Prediction of the MDS-UPDRS-III combining information of the user models based on GMM-UBM. \textbf{r:} Pearson's correlation coefficient, $\rho$: Spearman's correlation coefficient. MAE: median absolute error. }
\resizebox{\linewidth}{!}{
\begin{tabular}{lccccc}
\hline
                   & $\rho$     & p-val                       & \textbf{r}   & p-val                      & MAE  \\
                   \hline
Fusion of features & 0.634 & $\ll10^{-10}$ & 0.516 & $\ll10^{-7}$ & 10.5\\
\hline
\end{tabular}}
\label{tab:results2}
\end{table}

\begin{figure}[!ht]
    \centering
    \includegraphics[width=0.8\linewidth]{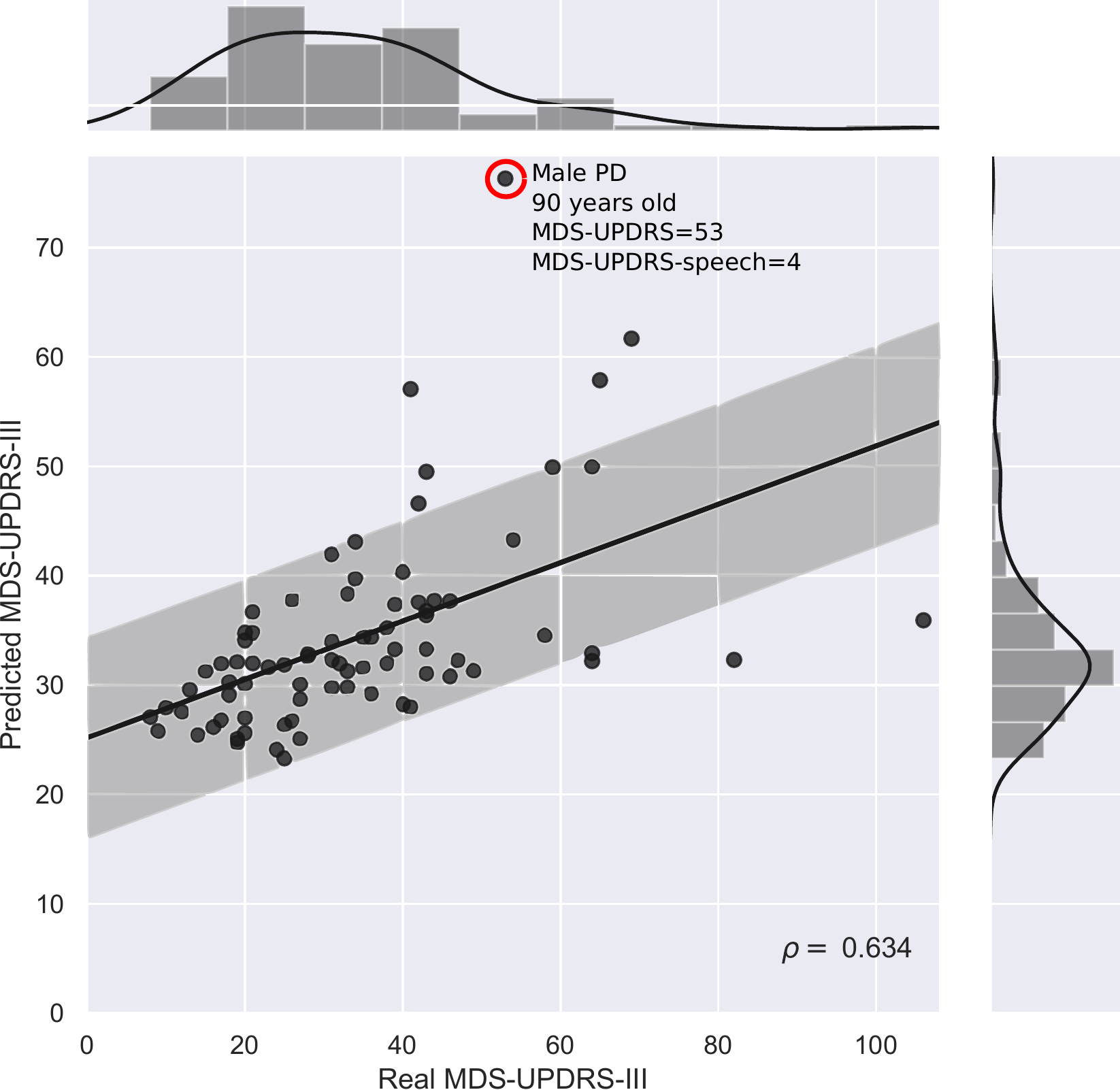}
    \caption{Prediction of the MDS-UPDRS-III score using multimodal user models based on GMM-UBM systems.}
    \label{fig:fusion}
\end{figure}

Figure~\ref{fig:fusion} shows the error in the prediction of the MDS-UPDRS-III score of the patients. Most of the patients are in initial or intermediate state of the disease (10$<$MDS-UPDRS-III$<$50), and they were predicted with the same distribution. The outlayer in the top of the figure corresponds to the eldest patient in the corpus. The patient has a score of 53, and it was predicted with a score of 78. Although the intermediate value of the MDS-UPDRS-III score of the patient, his MDS-UPDRS-speech item is 4, i.e., the patient was completely unable to speak, which highly affected his speech features.

\begin{figure}[!ht]
    \centering
    \includegraphics[width=0.75\linewidth]{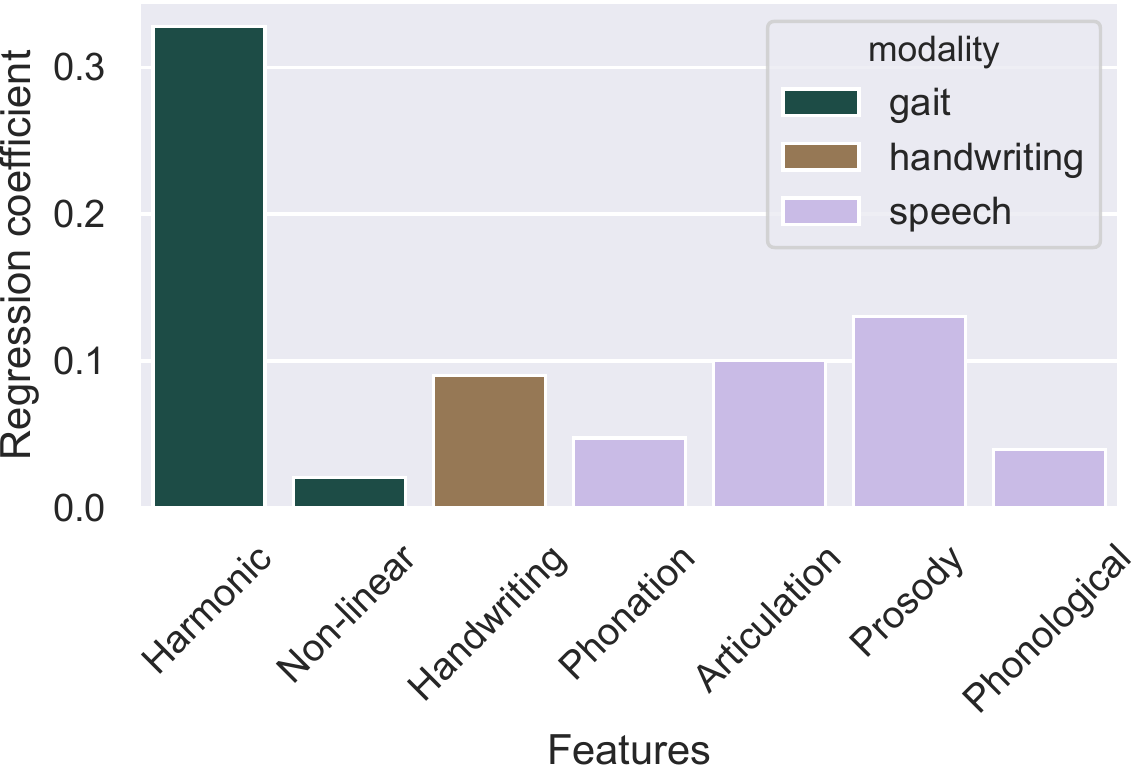}
    \caption{Contrbution of each feature set to the multimodal user model system.}
    \label{fig:fusion_bar}
\end{figure}

Figure~\ref{fig:fusion_bar} shows the contribution of each feature set to the multimodal user model. Each bar indicates the coefficient for the linear regression associated to each feature set. Harmonic features were the most important for the multimodal model, followed by prosody and articulation features, which have shown to be the most important features to evaluate the dysarthria associated to PD~\cite{vasquez2017gcca}. Handwriting features were less important than expected; however, this fact can be explained because the extracted features are based on a standard kinematic analysis that might not be completely related to the symptoms associated with PD. The results for handwriting could be improved with a feature set more related to the handwriting impairments of the patients, like those based on a neuromotor analysis~\cite{impedovo2019velocity}.

\section{Conclusion}
\label{sec:conclusion}
The present study compared user models based on GMM-UBM and i-vectors to evaluate the neurological state of PD patients using information from speech, handwriting, and gait. Different features were extracted from each bio-signal to model different dimensions of PD symptoms. Gait features were the most accurate to model the general neurological state of the patients, however, the combination of different bio-signals improved the correlation of the proposed method. In addition, user models based on GMM-UBM were more accurate than those based on i-vectors. Better results could be obtained with the i-vector system if more training data from HC subjects were available to create the reference model, especially for handwriting and gait. Further studies will consider additional features to model other aspects of PD symptoms, especially from handwriting signals. At the same time, additional models based on representation learning, and additional fusion methods can be considered to evaluate the neurological state of the patients.

\section{Acknowledgments}
\label{sec:ack}
This project received funding from the EU Horizon 2020 research and innovation programme under the Marie Sklodowska-Curie Grant Agreement No. 766287, and from CODI from University of Antioquia by Grant No.  2017--15530.

\vfill\pagebreak

\bibliographystyle{IEEEbib}
\bibliography{strings,refs}

\end{document}